\documentclass[pre,twocolumn,showpacs]{revtex4-1}
\usepackage{graphicx,amsmath,hyperref}
\newcommand{\bea}{\begin{eqnarray}}
\newcommand{\eea}{\end{eqnarray}}
\newcommand{\be}{\begin{equation}}
\newcommand{\ee}{\end{equation}}
\begin{document}
\def\fr{\frac}
\def\th{\theta}
\def\ph{\phi}
\def\l{\label}
\def\eps{\epsilon}
\def\o{\omega}
\def\eps{\epsilon}
\def\bef{\begin{figure}[h!]}
\def\eef{\end{figure}}

\title{Quasistationarity in a long-range interacting model of
particles moving on a sphere}
\author{Shamik Gupta$^1$}
\email{shamikg1@gmail.com}
\author{David Mukamel$^2$}
\email{david.mukamel@weizmann.ac.il}
\affiliation{$^1$Laboratoire de Physique Th\'{e}orique et Mod\`{e}les Statistiques,
UMR 8626, Universit\'{e} Paris-Sud 11 and CNRS,
B\^{a}timent 100, Orsay F-91405, France\\
$^2$Department of Physics of Complex Systems, Weizmann Institute of Science, Rehovot
76100, Israel}
\begin{abstract}
We consider a long-range interacting system of $N$ particles moving on a spherical surface under an
attractive Heisenberg-like interaction of infinite range, and evolving under deterministic Hamilton
dynamics. The system may also be viewed as one of globally coupled Heisenberg spins. In equilibrium,
the system has a continuous phase transition from a low-energy magnetized phase, in which the
particles are clustered on the spherical surface, to a high-energy homogeneous phase.
The dynamical behavior of the model is studied analytically by analyzing the Vlasov equation for the
evolution of the single-particle distribution, and numerically by direct simulations. The model is found to exhibit
long lived non-magnetized quasistationary states (QSSs) which in the thermodynamic limit are dynamically stable
within an energy range where the equilibrium state is magnetized. For
finite $N$, these states relax to
equilibrium over a time that increases algebraically with $N$. In the dynamically unstable regime,
non-magnetized states relax fast to equilibrium over a time that scales
as $\log N$. These features are retained
in presence of a global anisotropy in the magnetization.
\end{abstract}
\date{\today}
\maketitle
\section{Introduction}
Long-range interacting systems abound in nature
\cite{review1,review2,review3,review4,review5,review6}. In these
systems, the interparticle potential in $d$ dimensions decays at large
separation, $r$, as $1/r^{\alpha}$, with $0 \le \alpha \le d$. Some common examples
are self-gravitating systems \cite{Jeans:1919,Paddy:1990}, non-neutral plasmas \cite{Nicholson:1992}, two-dimensional geophysical
vortices \cite{Chavanis:2002}, dipolar ferroelectrics and ferromagnets
\cite{Landau:1960}, and many others. An important feature resulting from
long-range interactions is the property of non-additivity, whereby thermodynamic quantities scale superlinearly
with the system size. This results in equilibrium properties which
are generically not observed in short-range systems, e.g., a negative microcanonical
specific heat \cite{Lynden-Bell:1968,Thirring:1970}, inequivalence of
statistical ensembles \cite{Barre:2001,Mukamel:2005}; see
\cite{Bouchet:2005} for a discussion on classification of ensemble
inequivalence in long-range systems.

Non-additivity also has important consequences on the dynamical
properties, manifesting in broken
ergodicity \cite{Mukamel:2005,Borgonovi:2004,Borgonovi:2006,Bouchet:2008} and intriguingly slow
relaxation towards Boltzmann-Gibbs (BG) equilibrium \cite{Mukamel:2005, Chavanis:2002,Ruffo:1995,Yamaguchi:2004,Campa:2007,Joyce:2010}.
One of the first demonstrations of such slow relaxation has been in the
context of globular clusters for which dynamical evolution from a
non-steady initial to a quasistationary state has been shown to be as
long as 20 to 30 million years \cite{Henon:1964}.
A paradigmatic toy model that has been employed over the years for much
theoretical and numerical analysis of slow
relaxation in long-range interacting systems is the so-called Hamiltonian
mean-field (HMF) model. The model comprises globally
coupled particles moving on a unit circle and interacting via an
attractive $XY$-like interaction. The system evolves under
deterministic Hamilton dynamics. In this model, it has been found that for a wide
class of initial states, which have been termed quasistationary states
(QSSs), the relaxation time to equilibrium diverges with
the system size \cite{Ruffo:1995}. Moreover, for some energies, generic initial
states relax on a fast timescale of order one to such long-lived QSSs \cite{Yamaguchi:2004,Jain:2007}. As a consequence, such
systems
in the thermodynamic limit never attain the BG equilibrium
but remains trapped in the QSSs. Over the years, there have been several
theoretical studies aimed at explaining the observed features of QSSs in
the HMF model \cite{Antoniazzi:2007,Levin:2013,Ettoumi:2013}. Besides this model, QSSs have been observed in
many physical systems including single-pass high-gain free electron
laser \cite{Barre:2004} 
and two-dimensional electron plasma trapped in magnetic field
\cite{Kawahara:2006}.

In order to probe the ubiquity of quasistationary behavior, various
extensions of the HMF model have been introduced and analyzed over the
years. For example, it has been demonstrated that while anisotropic
versions of the HMF model do exhibit QSSs \cite{Jain:2007}, introducing
stochastic processes into the dynamics tends to destroy QSSs leading to
non-divergent relaxation times
\cite{Baldovin:2006,Baldovin:20061,Baldovin:2009,Gupta:2010,Gupta:20101,Chavanis:2011}.  A very
interesting generalization of the model to that of particles moving on
the surface of a sphere rather than on a circle has recently been introduced
\cite{Nobre:2003, Nobre:2004}. Here, the model is defined on a larger
phase space with each particle characterized by two positional degrees
of freedom rather than one as is the case for the HMF model. In this
model, the particles interact via an attractive Heisenberg-like
interaction of infinite range. The model may also be viewed as one of
globally coupled Heisenberg spins. The dynamics of the system follows
deterministic Hamilton equations of motion. In equilibrium, the system
exhibits a continuous phase transition from a low-energy magnetized
phase in which the particles are clustered on the spherical surface,
across critical threshold energy $\eps_c$, to a high-energy
homogeneous phase. Numerical studies of the model have shown that it
exhibits a number of quasistationary states, with relaxation times which
diverge with the system size \cite{Nobre:2003, Nobre:2004}. It would be
interesting to study analytically the relaxation processes in this higher dimensional model and trace the origin of its slow dynamics.

In this paper, we study analytically the relaxation dynamics of the model
of particles moving on a spherical surface, within the framework of the
Vlasov equation. In the limit $N \to \infty$, this equation describes the time evolution of the single-particle phase space
distribution. We show that there is an energy range $\epsilon^\star <
\epsilon < \epsilon_c$ where the BG equilibrium state is magnetized, in
which non-magnetized quasistationary states exist. This is manifested by the fact that these states are linearly stable stationary solutions of the Vlasov equation. For finite
$N$, however, such states relax to the equilibrium 
magnetized state over a time that scales algebraically with $N$. For energies
below $\eps^\star$, non-magnetized states, though stationary, are
linearly unstable under the Vlasov dynamics. Consequently, the system
exhibits a fast relaxation out of such initial states. These features remain unaltered on adding a term to the
Hamiltonian accounting for a global anisotropy in the
magnetization. 

The paper is organized as follows. In section \ref{isotropic}, we
describe the model of study, and analyze the Vlasov
equation to examine the relaxation dynamics. In particular, we show the
existence of non-magnetized QSSs in a specific energy range within the
thermodynamically stable magnetized phase. In section \ref{anisotropy}, we treat the case of an
additional global anisotropy in the Hamiltonian, and show similarities
in the relaxation dynamics when compared with the bare model. 
Finally, we draw our conclusions in section \ref{conclusions}.

\section{Isotropic Heisenberg mean-field model}
\l{isotropic}
Consider a system of $N$ interacting particles moving on the surface of a unit
sphere. The generalized coordinates of the $i$-th particle are the spherical polar angles $\th_i \in [0,\pi]$ and $\ph_i
\in [0,2\pi]$, while the corresponding generalized momenta are 
$p_{\th_i}$ and $p_{\ph_i}$. The Hamiltonian of the system is
given by 
\bea
H=\fr{1}{2}\sum_{i=1}^N \Big(p_{\th_i}^2+\fr{p_{\ph_i}^2}{\sin^2
\th_i}\Big)+\fr{1}{2N}\sum_{i,j=1}^N\Big[1-{\bf S}_i\cdot {\bf S}_j\Big].
\l{H}
\eea
Here, ${\bf S}_i$ is the vector pointing from the center to the position of the $i$-th particle on the sphere, and 
has the Cartesian components $(S_{ix},S_{iy},S_{iz})=(\sin
\th_i \cos \ph_i,\sin \th_i \sin \ph_i, \cos \th_i)$. The term involving
the double sum in Eq. \eqref{H}
describes the mean-field interaction between the particles. The
prefactor $1/N$ in the double sum makes the energy extensive, in
accordance with the Kac prescription \cite{Kac:1963}. Nevertheless, the
system is non-additive. This means that dividing the system into
macroscopic subsystems and summing over their thermodynamic variables
such as energy or entropy does not yield the corresponding variables of
the whole system. Regarding the vector ${\bf S}_i$ as the classical Heisenberg spin
vector of unit length, the interaction term has a form similar to that
in a mean-field Heisenberg model of magnetism. However, unlike the latter case, the
Poisson bracket between the components of ${\bf S}_i$'s in our model is
zero.

The interaction term in \eqref{H} tries to cluster the particles, and is
in competition with the kinetic energy term (the term involving
$p_{\th_i}$ and $p_{\ph_i}$) which has the opposite
effect.  The degree of clustering is conveniently measured by the
``magnetization" vector ${\bf m}=(m_x,m_y,m_z)=(1/N)\sum_{i=1}^N {\bf
S}_i$. In the BG equilibrium state, the system exhibits a
continuous phase transition at the critical energy density
$\eps_c=5/6$, between a low-energy clustered (``magnetized") phase in
which the particles are close together on the sphere, and a high-energy
homogeneous (``non-magnetized") phase in which the particles are uniformly distributed on
the sphere \cite{Nobre:2003,Nobre:2004}. As a function of the energy,
the magnitude of ${\bf m}$, given by $m=\sqrt{m_x^2+m_y^2+m_z^2}$, decreases continuously from unity at zero energy density to zero at $\eps_c$, and
remains zero at higher energies.

The time evolution of the system \eqref{H} follows the Hamilton
equations of motion, which for the $i$-th particle are given by
\bea
&&\fr{d\th_i}{dt}=p_{\th_i}, \l{eqnsofmotion1} \\
&&\fr{d \ph_i}{dt}=\fr{p_{\ph_i}}{\sin^2 \th_i}, \l{eqnsofmotion2} \\
&&\fr{dp_{\th_i}}{dt}=\fr{p^2_{\ph_i} \cos \th_i}{\sin ^3\th_i}\nonumber
\\
&&+m_x \cos \th_i \cos \ph_i+m_y \cos \th_i \sin \ph_i-m_z \sin
\th_i, 
\l{eqnsofmotion3} \\
&&\fr{dp_{\ph_i}}{dt}=-m_x \sin \th_i\sin \ph_i+m_y \sin \th_i \cos
\ph_i.\l{eqnsofmotion4} 
\eea
The dynamics conserves the total energy and the total momentum.

Here, we study how does the system starting far from
equilibrium and evolving under the dynamics, Eqs.
\eqref{eqnsofmotion1}-\eqref{eqnsofmotion4}, relax to the equilibrium state. To this end, we now derive
the Vlasov equation for the evolution of the phase space density. It is
known that for a mean-field system such as ours, this equation faithfully describes the $N$-particle
dynamics for finite time in the limit $N \to \infty$ \cite{Braun:1977}. In our
case, we conveniently study the dynamics by analyzing the motion of a
single particle in the four-dimensional phase space of its canonical
coordinates $(\th,\ph,p_\th,p_\ph)$, due to the mean-field produced by
its interaction with all the other particles. Let
$f(\th,\ph,p_\th,p_\ph,t)$ be the probability density in this
single-particle phase space, such that $f(\th,\ph,p_\th,p_\ph,t)d\th
d\ph dp_\th dp_\ph$ gives the probability at time $t$ to find the
particle with its generalized coordinates in $(\th,\th+d\th)$ and $(\ph,\ph+d\ph)$,
and the corresponding momenta in $(p_\th,p_\th+dp_\th)$ and
$(p_\ph,p_\ph+dp_\ph)$. Noting that the 
``velocity field" $(d\th/dt,d\ph/dt,dp_\th/dt,dp_\ph/dt)$ in the
phase space is divergence-free, conservation of probability 
implies that the total time derivative of $f$ vanishes:
\be 
\frac{df}{dt}=\frac{\partial f}{\partial t}+\fr{d\th}{dt}\frac{\partial
f}{\partial \th}+\fr{d\ph}{dt}\frac{\partial
f}{\partial \ph}+\fr{dp_\th}{dt}\frac{\partial
f}{\partial p_\th}+\fr{dp_\ph}{dt}\frac{\partial
f}{\partial p_\ph}=0.
\ee 
Using Eqs. \eqref{eqnsofmotion1}-\eqref{eqnsofmotion4} in the above equation, we get the Vlasov equation for time
evolution of $f(\th,\ph,p_\th,p_\ph,t)$ as
\begin{widetext}
\bea
&&\fr{\partial f}{\partial t}+p_\th\fr{\partial f}{\partial
\th}+\fr{p_\ph}{\sin^2\th}\fr{\partial f}{\partial \ph}+\Big(\fr{p^2_{\ph}
\cos \th}{\sin ^3\th}+m_x \cos
\th \cos \ph+m_y \cos \th \sin \ph-m_z \sin \th\Big)\fr{\partial f}{\partial
p_\th}\nonumber \\
&&+(-m_x \sin \th \sin \ph+m_y \sin \th \cos \ph)\fr{\partial
f}{\partial p_\ph}=0;
\l{Vlasov}\\
&&(m_x,m_y,m_z)=\int d\th d\ph dp_\th dp_\ph
~(\sin \th\cos \ph,\sin \th \sin \ph,\cos
\th)f.
\eea
\end{widetext}

Now, it is easily verified that any distribution
$f^{(0)}(\th,\ph,p_\th,p_\ph)=\Phi(e(\th,\ph,p_\th,p_\ph))$, with arbitrary
function $\Phi$, and $e$ being the single-particle energy,
\bea
&&e(\th,\ph,p_\th,p_\ph)=\fr{1}{2}\Big(p_{\th}^2+\fr{p_{\ph}^2}{\sin^2
\th}\Big)\nonumber \\
&&-m_x \sin \th \cos \ph-m_y \sin \th \sin \ph-m_z \cos \th,
\l{singlee}
\eea
is stationary under the Vlasov dynamics \eqref{Vlasov}. The magnetization
components $m_x,m_y,m_z$ are determined self-consistently. As a specific
example, consider a stationary state which is non-magnetized, that is,
$m_x=m_y=m_z=0$, and $f^{(0)}(\th,\ph,p_\th,p_\ph)$ is given by
\be
f^{(0)}(\th,\ph,p_\th,p_\ph)=\left\{ 
\begin{array}{ll}
               &\fr{1}{2\pi}\fr{1}{\pi}A \mbox{~~if~}
             \fr{1}{2}\Big(p_{\th}^2+\fr{p_{\ph}^2}{\sin^2
               \th} \Big)<E; \\
               &\th \in [0,\pi],\ph \in [0,2\pi],
               A,E\ge 0, \\
               &0,  \mbox{~~~~~~~~otherwise}.
               \end{array}
        \right. \\
\l{f0}        
\ee
The state \eqref{f0} is a straightforward generalization of the
well-studied water-bag initial condition for the HMF model
\cite{review3,Note:Tsallis}. The parameters $A$ and $E$ are related through the normalization
condition, 
\be
\int_{0}^{\pi}d\th\int_{0}^{2\pi}d\ph
\int_{\Omega}dp_\th
dp_{\ph}f^{(0)}=1,
\l{int1}
\ee
where the integration over $p_\th$ and $p_\ph$ is over the domain
$\Omega$ defined as
\be
\Omega=\Theta\Big(2E-p_{\th}^2-\fr{p_{\ph}^2}{\sin^2\th}\Big),
\ee
with $\Theta(x)$ denoting the unit step function. Performing the
integration in Eq. \eqref{int1}, we get
\be
E=\fr{1}{4A}.
\l{EA-relation}
\ee
The conserved energy density $\eps$, given by
\bea
\eps&=&\fr{1}{2}+\int_{0}^{\pi}d\th\int_{0}^{2\pi}d\ph
\int_\Omega dp_\th
dp_{\ph}\fr{1}{2}\Big(p_{\th}^2+\fr{p_{\ph}^2}{\sin^2\th}\Big)f^{(0)},\nonumber
\\
\eea
is related to the parameter $E$ as
\be
\eps=\fr{1}{2}+\fr{E}{2}.
\l{energy}
\ee

We now examine the stability of the stationary state
 \eqref{f0} under the dynamics \eqref{Vlasov}. In
particular, we will study linear stability. To this end, consider small perturbation around the
state $f^{(0)}$, so that the corresponding state may be expanded as
\be
f(\th,\ph,p_\th,p_\ph,t)=f^{(0)}+\delta f(\th,\ph,p_\th,p_\ph,t);
~~\delta f \ll 1.
\l{fexpansion}
\ee
Since both $f$ and $f^{(0)}$ are normalized, we have
\be
\int d\th d\ph dp_\th dp_\ph \delta f(\th,\ph,p_\th,p_\ph,t)=0.
\l{deltaf-norm}
\ee
Using Eq. \eqref{fexpansion} in Eq. \eqref{Vlasov}, and keeping terms to linear order in
$\delta f$, we get
\bea
&&\fr{\partial \delta f}{\partial t}+p_\th\fr{\partial \delta
f}{\partial \th}+\fr{p_\ph}{\sin^2\th}\fr{\partial \delta f}{\partial \ph}+\fr{p^2_{\ph}
\cos \th}{\sin ^3\th}\fr{\partial \delta f}{\partial p_\th}\nonumber \\
&&+(\widetilde{m}_x \cos \th \cos \ph+\widetilde{m}_y \cos \th \sin
\ph-\widetilde{m}_z \sin \th)\fr{\partial f^{(0)}}{\partial p_\th}\nonumber \\
&&+(-\widetilde{m}_x \sin \th \sin \ph+\widetilde{m}_y \sin \th \cos
\ph)\fr{\partial f^{(0)}}{\partial p_\ph}=0;
\l{linearVlasov} \\
&&(\widetilde{m}_x,\widetilde{m}_y,\widetilde{m}_z)\nonumber \\
&&=\int d\th d\ph dp_\th dp_\ph
~(\sin \th\cos \ph,\sin \th \sin \ph,\cos
\th)\delta f. 
\eea
The linearized dynamics at long times is expected to be dominated
by the mode corresponding to the largest eigenfrequency $\o$ of the
linearized equation \eqref{linearVlasov}, so that we may write
\be
\delta f(\th,\ph,p_\th,p_\ph,t)=\widetilde{\delta f}(\th,\ph,p_\th,p_\ph,\o)e^{i\o t}.
\ee

When the state \eqref{f0} is linearly unstable, the system gets
magnetized, which
due to the complete isotropy of the Hamiltonian \eqref{H} may be taken
to be along the $z$ direction without any loss of generality. This implies (and is
implied by) a form of perturbation which is uniform in $\ph$:
\be
\widetilde{\delta f}(\th,\ph,p_\th,p_\ph,\o)=\fr{1}{2\pi}g(\th,p_\th,p_\ph,\o).
\ee
In this case, it follows from Eq. \eqref{eqnsofmotion4} that for all
$i$, 
\be
p_{\ph_i}= {\rm constant},
\ee
equal to its initial value, so that 
\be
\int d\th dp_\th g(\th,p_\th,p_\ph,\o)=0.
\l{marginal-condition}
\ee

Using the above arguments and Eq. \eqref{linearVlasov}, it follows that
corresponding to the neutral mode $\o=0$, the quantity $g_0=g(\th,p_\th,p_\ph,0)$ satisfies
\bea
&&p_\th \fr{\partial g_0}{\partial \th}+\fr{p^2_{\ph}
\cos \th}{\sin^3\th}\fr{\partial g_0}{\partial
p_\th}\nonumber \\
&&-\fr{A}{2\pi^2}\Big[\delta(p_\th+p_0(\th,p_\ph))-\delta(p_\th-p_0(\th,p_\ph))\Big]\widetilde{m}_{z,0}
\sin \th=0, \nonumber \\
\l{linearVlasov-01}
\eea
where $\delta$ is the Dirac Delta function, while 
\be
\widetilde{m}_{z,0}=2\pi\int_0^{\pi} d\th \cos \th 
\int_{\Omega}
dp_\th 
dp_\ph g_0, 
\l{mz-defn}
\ee
and 
\be
p_0(\th,p_\ph)=\sqrt{2E-p_\ph^2/\sin^2\th}.
\ee
In arriving at Eq. \eqref{linearVlasov-01}, we have used Eq. \eqref{f0} to obtain
\be
\fr{\partial f^{(0)}}{\partial
p_\th}=\fr{A}{2\pi^2}\Big[\delta(p_\th+p_0(\th,p_\ph))-\delta(p_\th-p_0(\th,p_\ph))\Big].
\ee
From Eq. \eqref{linearVlasov-01}, we see that $g_0$ has the property
\be
g_0(\th,p_\th,p_\ph)=-g_0(\pi-\th,p_\th,p_\ph).
\l{symmetry}
\ee

We now solve Eq. \eqref{linearVlasov-01} for $g_0$. To proceed,
consider a solution of the form
\bea
g_0(\th,p_\th,p_\ph)&=&
\Big[a(\th,p_\ph)\delta(p_\th+p_0(\th,p_\ph))\nonumber \\
&&+b(\th,p_\ph)\delta(p_\th-p_0(\th,p_\ph))\Big].
\l{g-expansion}
\eea
Equation \eqref{marginal-condition} then implies that
\be
\int d\th \Big(a(\th,p_\ph)+b(\th,p_\ph)\Big)=0.
\l{ab-cond}
\ee
Also, Eq. \eqref{symmetry} implies that 
\be
a(\pi/2,p_\ph)=b(\pi/2,p_\ph)=0.
\l{symmetry-1}
\ee

Substituting Eq. \eqref{g-expansion} into Eq. \eqref{linearVlasov-01} gives
\bea
&&\fr{p^2_{\ph}
\cos
\th}{\sin^3\th}a\Big(1+\fr{p_\th}{p_0(\th,p_\ph)}\Big)\delta'(p_\th+p_0(\th,p_\ph))\nonumber
\\
&&-\Big(p_0(\th,p_\ph) \fr{\partial
a}{\partial
\th}+\fr{A}{2\pi^2}\widetilde{m}_{z,0}\sin\th\Big)\delta(p_\th+p_0(\th,p_\ph))=0,
\nonumber \\
\l{aeqn} \\
&&\fr{p^2_{\ph}
\cos
\th}{\sin^3\th}b\Big(1-\fr{p_\th}{p_0(\th,p_\ph)}\Big)\delta'(p_\th-p_0(\th,p_\ph))\nonumber
\\
&&+\Big(p_0(\th,p_\ph) \fr{\partial
b}{\partial
\th}+\fr{A}{2\pi^2}\widetilde{m}_{z,0}\sin\th\Big)\delta(p_\th-p_0(\th,p_\ph))=0.\nonumber
\\
\l{beqn} 
\eea
Using $x\delta'(x)=-\delta(x)$, the above equations give
\bea
&&-\Big(\frac{d(ap_0(\th,p_\ph))}{d\th}+\fr{A}{2\pi^2}\widetilde{m}_{z,0}\sin\th\Big)\delta(p_\th+p_0(\th,p_\ph))=0,
\nonumber \\\l{aeqn2} \\
&&\Big(\frac{d(bp_0(\th,p_\ph))}{d\th}+\fr{A}{2\pi^2}\widetilde{m}_{z,0}\sin\th\Big)\delta(p_\th-p_0(\th,p_\ph))=0.
\nonumber \\ \l{beqn2} 
\eea
We thus have
\bea
&&\fr{d a(\th,p_\ph)p_0(\th,p_\ph)}{d
\th}=\fr{db(\th,p_\ph)p_0(\th,p_\ph)}{d \th}=-\fr{A\widetilde{m}_{z,0}\sin
\th}{2\pi^2}. \nonumber \\
\l{ab-before-int}
\eea
Integrating between $\th'=\pi/2$ and $\th'=\th$, and using
Eq. \eqref{symmetry-1}, we get 
\bea
&&a(\th,p_\ph)=b(\th,p_\ph)=\fr{A\widetilde{m}_{z,0}\cos
\th}{2\pi^2p_0(\th,p_\ph)}.
\l{ab-1}
\eea
We therefore have from Eq. \eqref{g-expansion} the solution of the linearized
Vlasov equation \eqref{linearVlasov-01} as 
\bea
&&g_0(\th,p_\th,p_\ph)=\fr{A\widetilde{m}_{z,0}\cos
\th}{2\pi^2p_0(\th,p_\ph)}\nonumber \\
&&\times
\Big[\delta(p_\th+p_0(\th,p_\ph))+\delta(p_\th-p_0(\th,p_\ph))\Big]
\nonumber \\
&&=\fr{A\widetilde{m}_{z,0}\cos
\th}{\pi^2}\delta^2(2E-p_\th^2-p_\ph^2/\sin^2 \th),
\l{g-expansion-final}
\eea
which is invariant under rotation about the $z$-axis, as it should be.

Equations \eqref{mz-defn} and
\eqref{g-expansion-final} give a self-consistent equation for
$\widetilde{m}_{z,0}$, as follows:
\begin{widetext}
\bea
&&\widetilde{m}_{z,0}=2\pi \int_0^{\pi}d\th \cos \th \int_{-\sqrt{2E}\sin
\th}^{\sqrt{2E}\sin \th}
dp_\ph \int_{-\sqrt{2E-p^2_\ph/\sin^2\th}}^{\sqrt{2E-p^2_\ph/\sin^2\th}}
dp_\th g_0(\th,p_\th,p_\ph)
\nonumber \\
&&=\fr{A\widetilde{m}_{z,0}}{\pi^2}2\pi \int_0^{\pi} d\th \cos^2\th \int_{-\sqrt{2E}\sin
\th}^{\sqrt{2E}\sin
\th}dp_\ph\frac{dp_\ph}{\sqrt{2E-p_\ph^2/\sin^2\th}}\nonumber \\
&&=\frac{4A\widetilde{m}_{z,0}}{3}.
\eea
\end{widetext}

We therefore have the desired self-consistent equation for
$\widetilde{m}_{z,0}$:
\bea
\widetilde{m}_{z,0}\Big[1-\fr{4A}{3}\Big]=0.
\eea
Then, since $\widetilde{m}_{z,0}\ne 0$, the above equation is satisfied
with $A=A^\star$, where $A^*=3/4$.
Correspondingly, on using Eqs.
\eqref{EA-relation} and \eqref{energy}, we obtain the energy threshold for the linear stability of the state
\eqref{f0} as
\bea
\eps^*=\fr{2}{3}.
\eea
On the basis of our analysis, we thus conclude that in the energy range
$\eps^* < \eps <\eps_c$, the non-magnetized state (\ref{f0}) is linearly
stable, and is hence a QSS. In a finite system, a QSS eventually
relaxes to BG equilibrium on a timescale over which non-linear
correction terms should be added to the Vlasov equation \cite{review3}.
In the HMF model, numerical simulations \cite{Yamaguchi:2004} have shown
this timescale to grow with system size $N$ as $N^{\delta}; \delta
\simeq 1.7$. Recent extensive numerical studies suggest that in fact for
the HMF model, $\delta \simeq 2$ \cite{Figueiredo:2013}.

In order to verify the above prediction of QSSs in our model, we performed numerical
simulations of the dynamics by integrating the equations of
motion (\ref{eqnsofmotion1}), (\ref{eqnsofmotion2}),
(\ref{eqnsofmotion3}) and (\ref{eqnsofmotion4}) by using a fourth-order Runge-Kutta method with time step equal
to $10^{-2}$. For energies in between $\eps^*$ and $\eps_c$, the results
shown in Fig. \ref{N2-K0} indeed show that consistent with our
predictions, the initial state (\ref{f0}) is a QSS, relaxing to BG
equilibrium over a very long timescale that grows algebraically with the
system size as $N^\delta$. The scaling collapse plot of Fig.
\ref{N2-K0}(b) suggests that $\delta \simeq 1.7$.  
For energies $\eps < \eps^*$, when the state (\ref{f0}) is unstable,
Fig. \ref{lnN-K0} illustrates that the system exhibits a fast relaxation
towards BG equilibrium over a timescale that grows with the system size
as $\log N$. 

\bef
\begin{center}
\includegraphics[width=80mm]{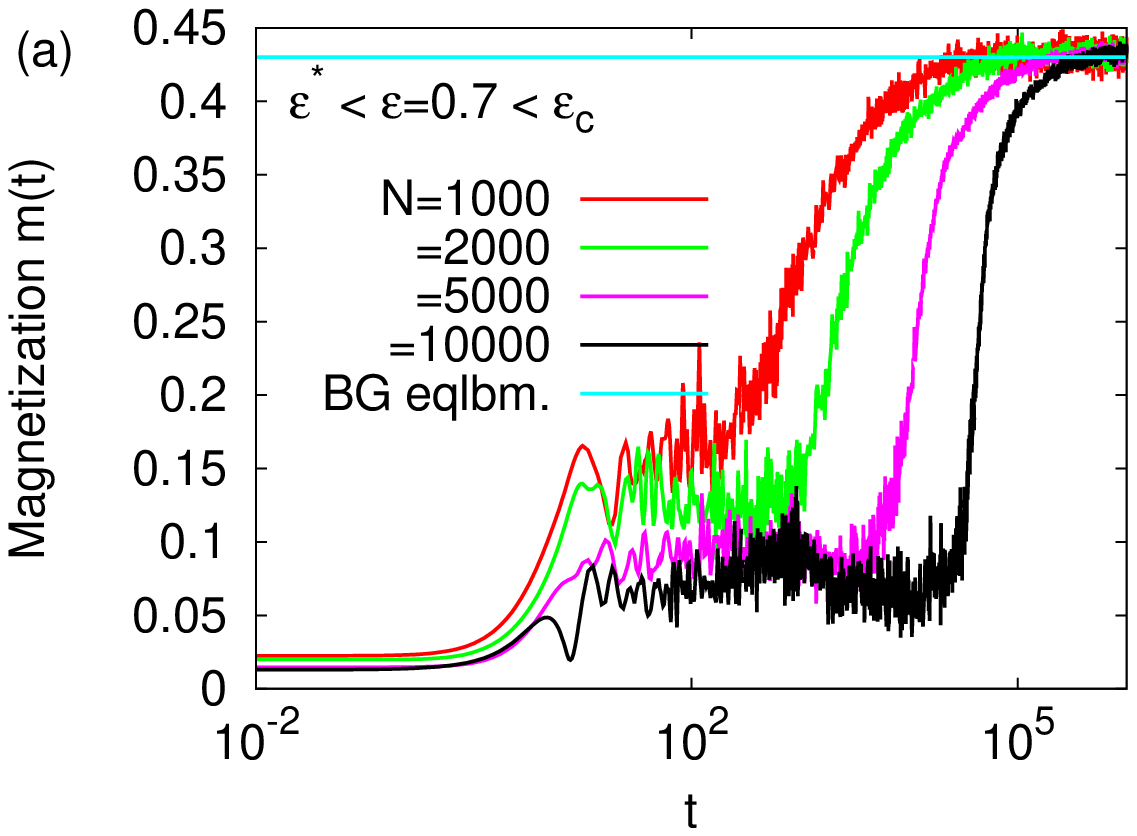}
\includegraphics[width=80mm]{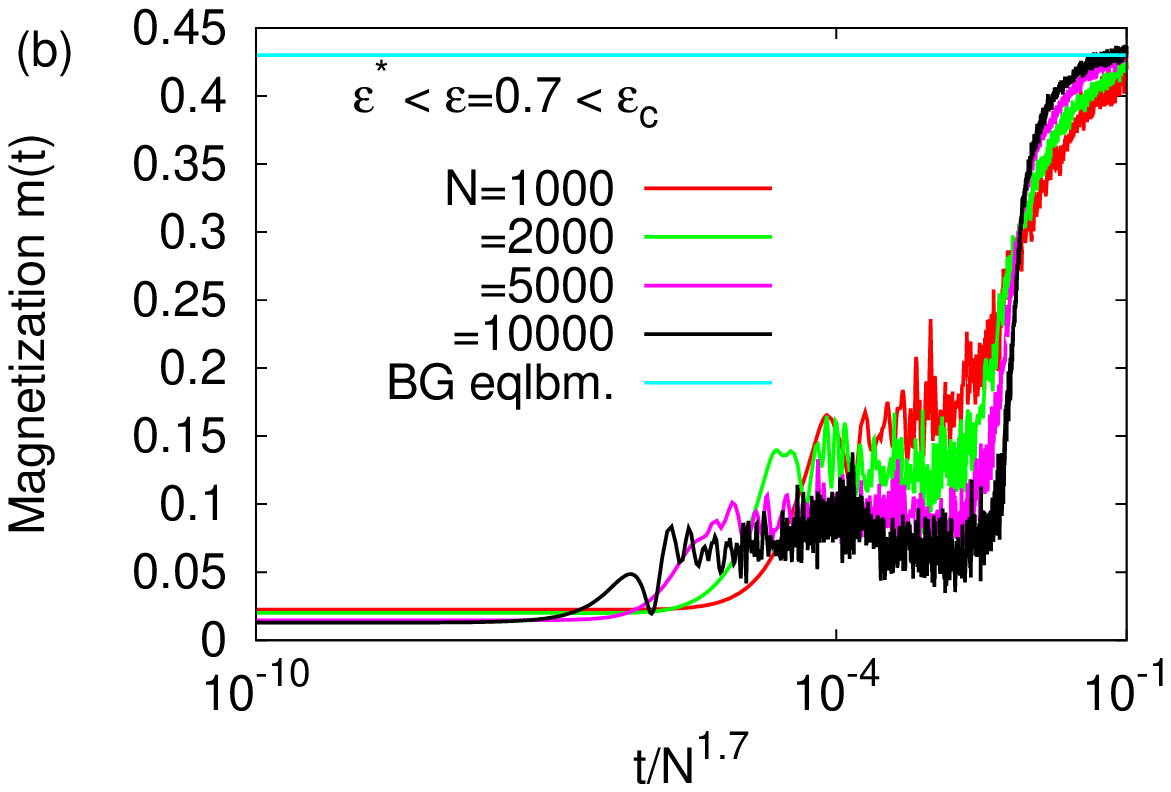}
\caption{For the model \eqref{H}, the figures show the magnetization $m(t)$ as a function of time
(a), and as a function of time scaled by $N^{1.7}$ (b) in the Vlasov-stable phase ($\eps^\star < \eps <
\eps_c$) for the model (\ref{H}) at energy density $\eps=0.7$. The
figures suggest the existence of a QSS with a lifetime scaling with the
system size as $N^{1.7}$. Data averaging varies between $5$ histories for
the largest system and
$10$ histories for the smallest one.}
\l{N2-K0}
\end{center}
\eef

\bef
\begin{center}
\includegraphics[width=80mm]{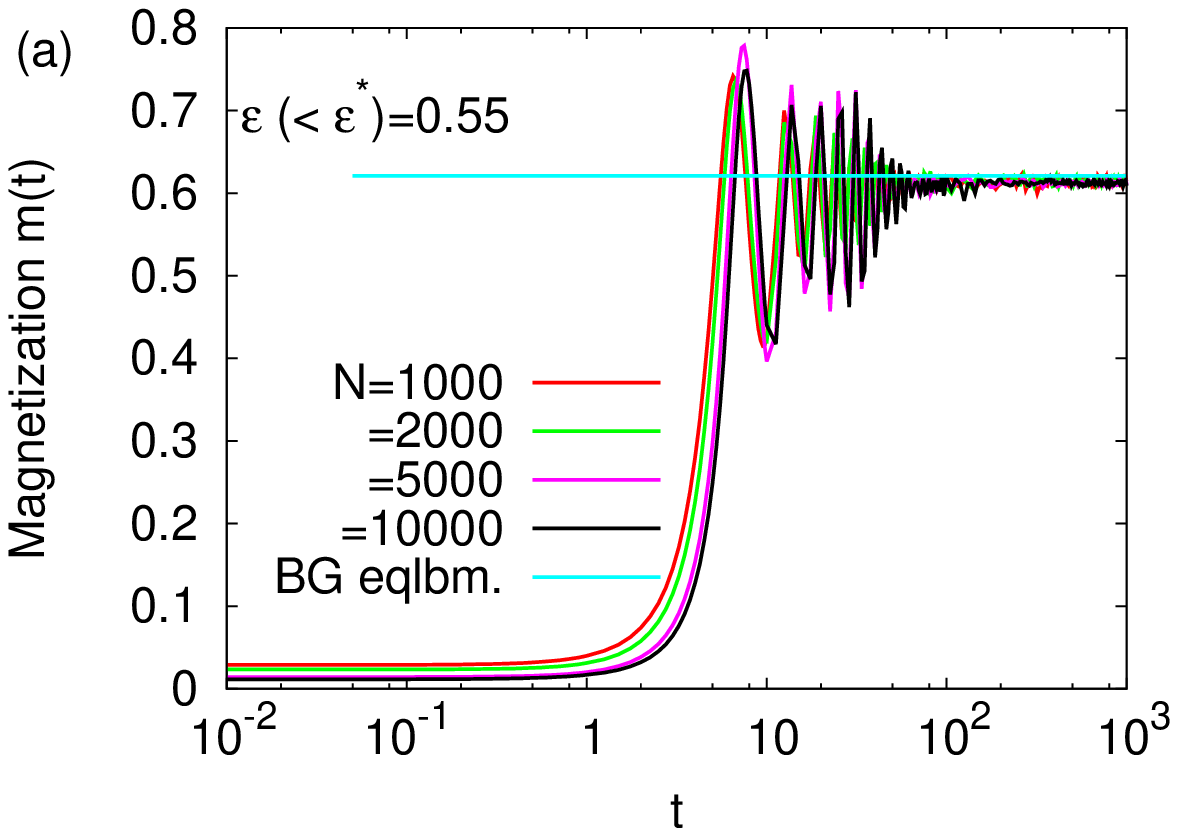}
\includegraphics[width=80mm]{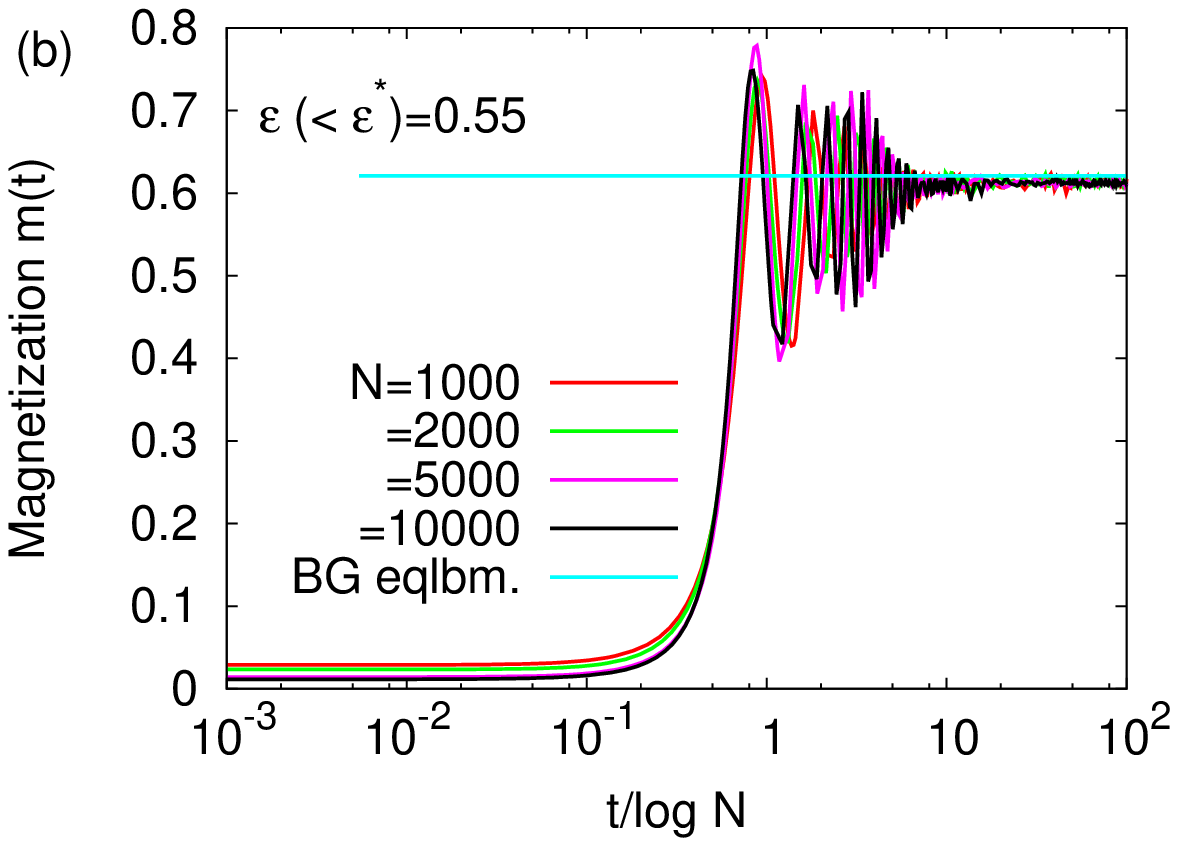}
\caption{For the model \eqref{H}, the figures show the magnetization $m(t)$ as a function of time
(a), and as a function of time scaled by the logarithm of the
system size $N$ (b) in the Vlasov-unstable phase ($\eps <
\eps^\star$) at energy density $\eps=0.55$. The
figures show a fast relaxation from the initial non-magnetized state
to BG equilibrium over a timescale $\sim \log N$. 
Data averaging varies between $5$ histories for the largest system and
$10$ histories for the smallest one.}
\l{lnN-K0}
\end{center}
\eef

In the next section, we modify the model \eqref{H} to include an
additional global anisotropy.
\section{Heisenberg mean-field model with additional global anisotropy}
\l{anisotropy}
In this section, we consider the system \eqref{H} with an additional
global anisotropy, and demonstrate the existence of QSSs, similar to the
bare model. The analysis is similar to
that in the previous section, and therefore, here we
briefly outline the main steps. The Hamiltonian of the system is given by
\bea
&&H=\fr{1}{2}\sum_{i=1}^N \Big(p_{\th_i}^2+\fr{p_{\ph_i}^2}{\sin^2
\th_i}\Big)+\fr{1}{2N}\sum_{i,j=1}^N\Big[1-{\bf S}_i\cdot {\bf
S}_j\Big]\nonumber \\
&&-\fr{K}{2N}(\sum_{i=1}^N S_{iz})^2,
\l{H-2}
\eea
where the last term gives the energy due to a global anisotropy in the
magnetization along the $z$ direction. For simplicity, we consider $K > 0$ for which the energy is minimized by ordering along the $z$-axis.
At the end of this section, we will comment on the case $K<0$.

Following standard
procedure (see, e.g., \cite{Jain:2007}), the equilibrium magnetization
$\langle m_z \rangle$ satisfies
\be
\langle m_z \rangle = \fr{\int d\th d\ph \cos \th \sin \th e^{\beta
(K+1) \langle m_z \rangle \cos \th}}{\int d\th d\ph \sin \th e^{\beta (K+1) \langle m_z
\rangle \cos \th}}.
\ee
Close to the critical point, expanding the above equation to leading
order in $\langle m_z \rangle$, we get
\bea
\langle m_z \rangle(2-\beta (K+1)\int_0^\pi d\th \sin \th \cos^2 \th)=0.
\eea
With $\langle m_z \rangle \ne 0$, we get the critical temperature
$\beta_c$ as
\be
\beta_c=\fr{3}{K+1}.
\l{betac-2}
\ee
The critical energy density is
\be
\eps_c(K)=\fr{1}{\beta_c}+\fr{1}{2}=\fr{5}{6}+\fr{K}{3}; K>0.
\l{epsc-2}
\ee

The Hamilton equations of motion for the model \eqref{H-2} are obtained
from Eqs. \eqref{eqnsofmotion1} - \eqref{eqnsofmotion4} by replacing
$m_z$ by $(K+1)m_z$. The Vlasov equation for the evolution of the single-particle
phase space density $f(\th,\ph,p_\th,p_\ph,t)$ may be derived as in the
previous section, and is given by Eq. \eqref{Vlasov} with $(K+1)m_z$
replacing $m_z$. 

Now, any distribution
$f^{(0)}(\th,\ph,p_\th,p_\ph)=\Phi(e(\th,\ph,p_\th,p_\ph))$, with arbitrary
function $\Phi$, and $e$ being the single-particle energy,
\bea
&&e(\th,\ph,p_\th,p_\ph)=\fr{1}{2}\Big(p_{\th}^2+\fr{p_{\ph}^2}{\sin^2
\th}\Big)\nonumber \\
&&-m_x \sin \th \cos \ph-m_y \sin \th \sin \ph-(K+1)m_z \cos
\th,\nonumber \\ 
\l{singleeagain-2}
\eea
is stationary under the Vlasov dynamics. In particular, the
non-magnetized state
\eqref{f0} represents a stationary solution of the Vlasov dynamics. 

Let us now study the linear stability of the state
\eqref{f0}. Noting that when the state is
unstable, the system for $K>0$ orders along the $z$ direction, 
we linearize the Vlasov equation about the state, by writing
\be
f(\th,\ph,p_\th,p_\ph,t)=f^{(0)}+\fr{1}{2\pi}g(\th,p_\th,p_\ph,\o)e^{i\o
t},
\ee
so that corresponding to the neutral mode $\o=0$, the quantity $g_0=g(\th,p_\th,p_\ph,0)$ satisfies
\bea
&&p_\th \fr{\partial g_0}{\partial \th}+\fr{p^2_{\ph}
\cos \th}{\sin^3\th}\fr{\partial g_0}{\partial
p_\th}\nonumber \\
&&-\fr{A}{2\pi^2}\Big[\delta(p_\th+p_0(\th,p_\ph))-\delta(p_\th-p_0(\th,p_\ph))\Big]\nonumber
\\
&&\times (K+1)\widetilde{m}_{z,0}
\sin \th=0. 
\l{linearVlasov-2}
\eea
The above equation is similar to Eq. \eqref{linearVlasov-01}, the
only difference being an extra constant factor $(K+1)$ in the term
involving $\widetilde{m}_{z,0}$. Consequently, the
analysis following Eq. \eqref{linearVlasov-01} may be similarly carried
out in the present case to get
\bea
&&a(\th,p_\ph)=b(\th,p_\ph)=\fr{A(K+1)\widetilde{m}_{z,0}\cos
\th}{2\pi^2p_0},
\l{ab-2}
\eea
which may be combined with Eqs.
\eqref{mz-defn} and \eqref{g-expansion} to obtain the following equation:
\be
\widetilde{m}_{z,0}\Big[1-(K+1)\fr{4A}{3}\Big]=0.
\l{Astareqn-2}
\ee
It then implies that 
 corresponding to the neutral mode, one has
$A=A^\star=3/(4(K+1))$, which together with
Eqs. \eqref{EA-relation} and
\eqref{energy} give the energy density $\eps^\star(K)$ corresponding to
neutral stability of the stationary state \eqref{f0} under the
linearized Vlasov dynamics:
\be
\eps^\star(K)=\fr{2}{3}+\fr{K}{6};
~~K>0.
\l{eps-star-2}
\ee

Compared to the bare model, we thus see that global anisotropy widens
the range of energy $\eps^*(K) < \eps <\eps_c(K)$ over which the
non-magnetized state (\ref{f0}) is linearly stable under the Vlasov
dynamics and is hence a QSS. For energies below $\eps^*(K)$, such a
state being linearly unstable exhibits a fast relaxation towards 
BG equilibrium over a timescale $\sim \log N$, while for
energies in the range $\eps^*(K) < \eps <\eps_c(K)$, it relaxes to BG
equilibrium only over a very long timescale growing with the system size
as $N^{1.7}$, see Figs. \ref{N2-K1} and \ref{lnN-K1}.

For $K<0$, the system will order in the $xy$-plane, and the anisotropy
term does not affect the energy. Indeed, an analysis along the same
lines as above shows that the energy thresholds $\eps^\star$ and
$\eps_c$ are equal to the corresponding values for the bare model
($K=0$), and non-magnetized QSSs exist in the range $\eps^\star < \eps <
\eps_c$ ($\eps^*=2/3, \eps_c=5/6$). 

\bef
\begin{center}
\includegraphics[width=80mm]{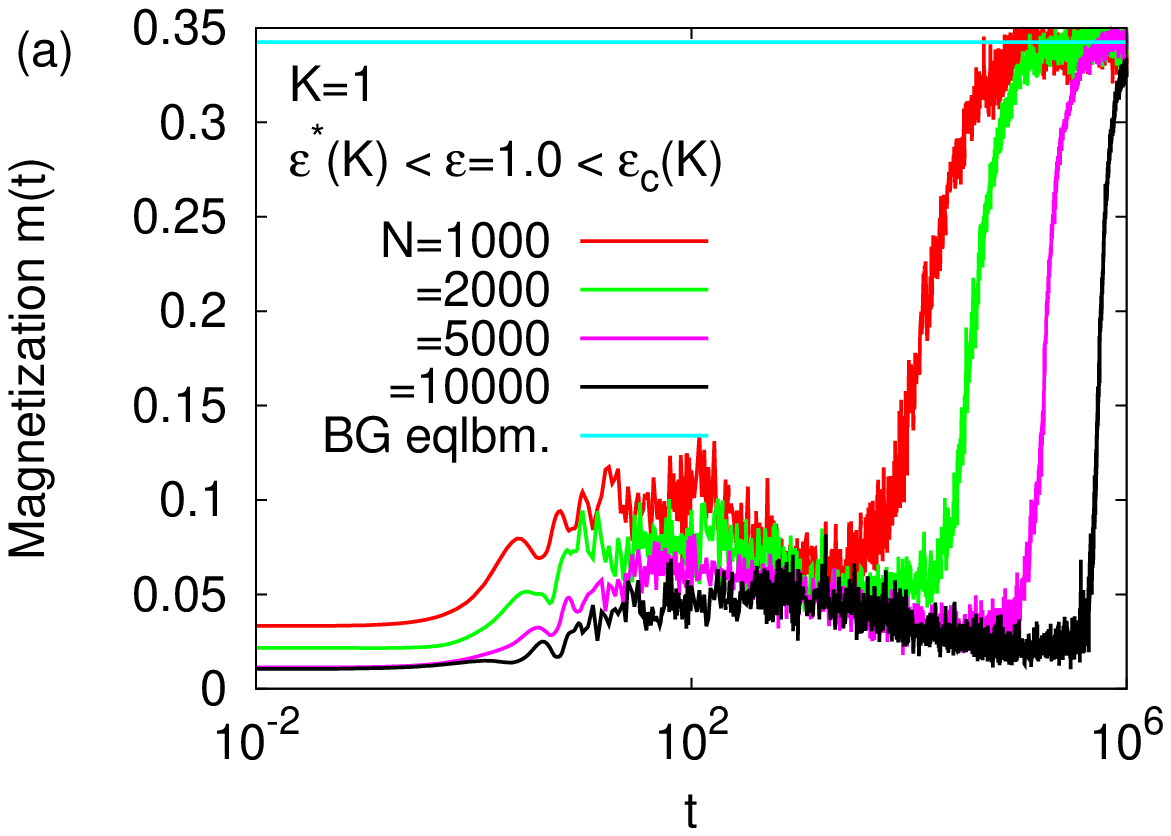}
\includegraphics[width=80mm]{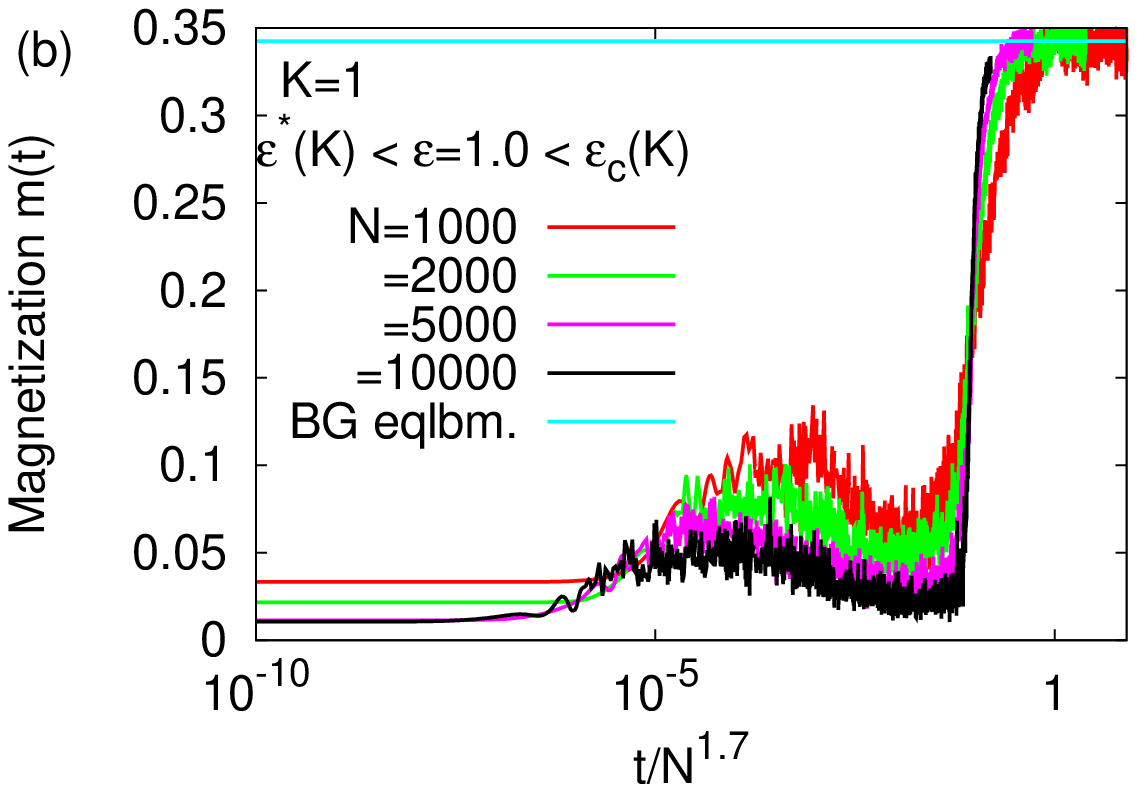}
\caption{For the model \eqref{H-2}, the figures show the magnetization $m(t)$ as a function of time
(a), and as a function of time scaled by $N^{1.7}$ (b) in the Vlasov-stable phase
($\eps^\star(K) < \eps <
\eps_c(K)$) for $K=1$ at energy density $\eps=1.0$. The
figures suggest the existence of a QSS with a lifetime scaling with the
system size as $N^{1.7}$. Data averaging varies between $6$ histories
for the largest system and
$10$ histories 
for the smallest one.}
\l{N2-K1}
\end{center}
\eef

\bef
\begin{center}
\includegraphics[width=80mm]{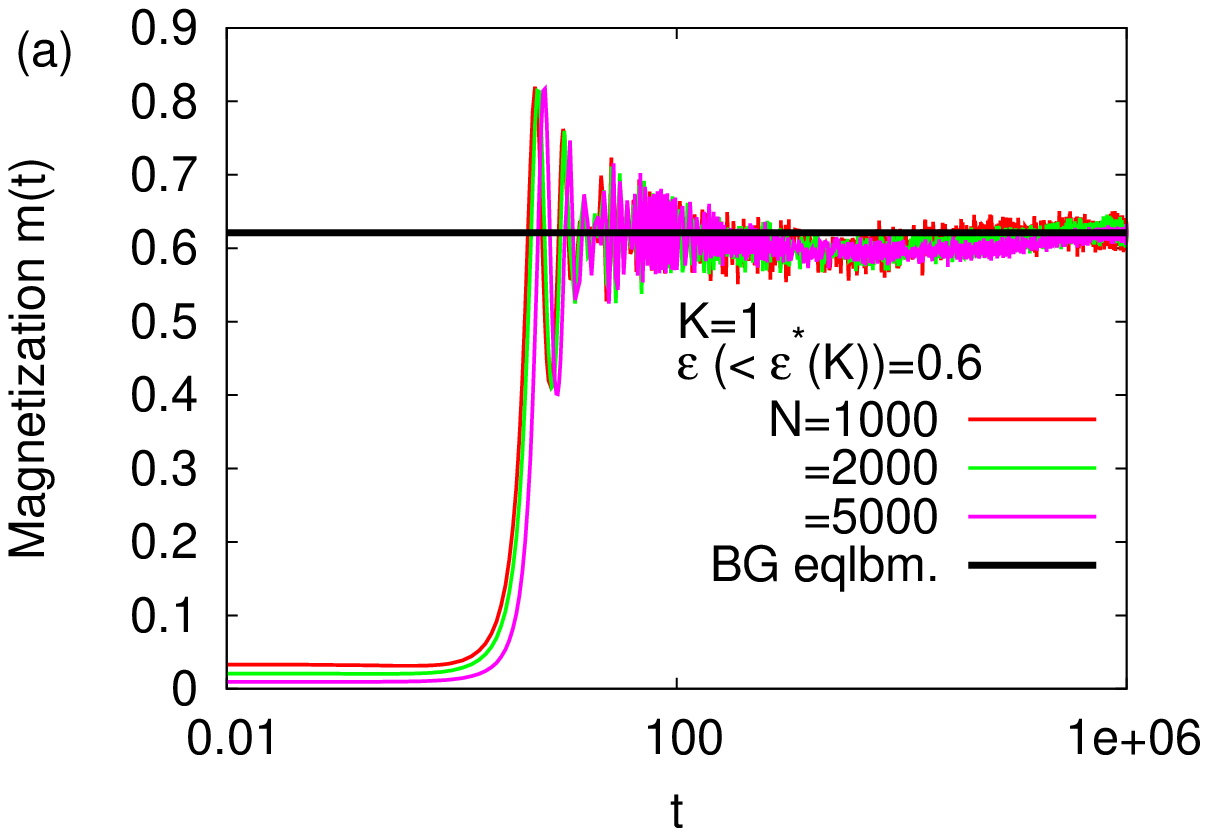}
\includegraphics[width=80mm]{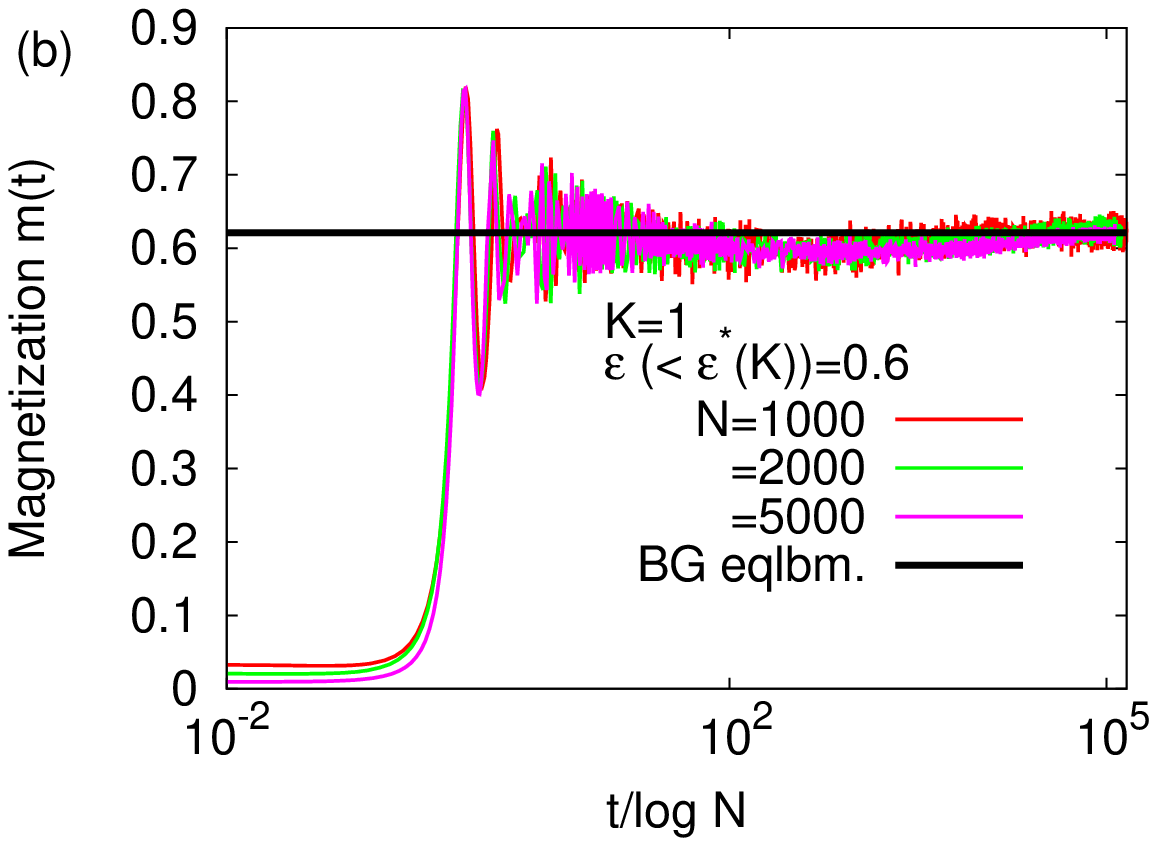}
\caption{For the model \eqref{H-2}, the figures show the magnetization $m(t)$ as a function of time
(a), and as a function of time scaled by the logarithm of the
system size $N$ (b) in the Vlasov-unstable phase ($\eps <
\eps^\star(K)$) for $K=1$ at energy density $\eps=0.6$. The
figures show a fast relaxation from the initial non-magnetized state
to BG equilibrium over a timescale $\sim \log N$. 
The data shown are for one realization of the
dynamics.}
\l{lnN-K1}
\end{center}
\eef

\section{Conclusions}
\l{conclusions}
In this work, we addressed the ubiquity of non-Boltzmann quasistationary
states (QSS) in long-range systems. This is done by analyzing the relaxation dynamics
of a system of $N$ particles moving on a
spherical surface under an attractive Heisenberg-like interaction of
infinite range, and
evolving under deterministic Hamilton dynamics. 
In equilibrium, the system exhibits a continuous phase transition from a
low-energy magnetized phase to a high-energy homogeneous
phase at the energy density $\eps_c=5/6$. In the limit of infinite $N$,
the dynamics of relaxation to equilibrium is described by the Vlasov
equation for the temporal evolution of the single-particle phase space
distribution. By linearizing the Vlasov equation about a stationary
non-magnetized state, our exact solution of the linearized equation
shows that within the thermodynamically stable magnetized phase, there
exists an energy range $\eps^\star < \eps < \eps_c$ over which
non-magnetized states occur as 
stable stationary solutions of the Vlasov dynamics, where $\eps^\star$
corresponds to linear stability threshold of the nonmagnetized states. This leads to the
formation of long-lived non-magnetized quasistationary states (QSSs),
with a lifetime that we demonstrate on the basis of numerical
simulations to be growing algebraically with the system size $N$. For
energies below $\eps^\star$, non-magnetized stationary states are
linearly unstable under the Vlasov dynamics, and thus, exhibit a fast
relaxation to equilibrium over a timescale growing with the system size
as $\log N$. These features remain unaltered on adding a term to the
Hamiltonian that accounts for a global anisotropy in the
magnetization. 
\section{Acknowledgements}
We acknowledge fruitful discussions with Freddy Bouchet, Or Cohen, Thierry
Dauxois, and Stefano
Ruffo. We are grateful to \'{E}cole Normale Sup\'{e}rieure de Lyon and Korea Institute for Advanced Study for
hospitality during our mutual visit when part of this work was done. 
S. G. acknowledges the CEFIPRA Project 4604-3 and the contract ANR-10-CEXC-010-01 for
support. We gratefully acknowledge the support of the Israel Science Foundation (ISF) and the Minerva Foundation with funding from the Federal German Ministry for Education and Research.


\begin{thebibliography}{99}
\bibitem{Note:Tsallis}Note that the initial condition referred to as
water-bag in Refs. \cite{Nobre:2003,Nobre:2004} is very different from
the state \eqref{f0}.
\end{thebibliography}
\end{document}